\documentclass[12pt,preprint]{aastex}

\usepackage{multirow}
\usepackage{hhline}
\usepackage{url}
\usepackage{amssymb}

\def\ltsima{$\; \buildrel < \over \sim \;$}
\def\gtsima{$\; \buildrel > \over \sim \;$}
\def\lsim{\lower.5ex\hbox{\ltsima}}
\def\gsim{\lower.5ex\hbox{\gtsima}}
\def\lapp{\ifmmode\stackrel{<}{_{\sim}}\else$\stackrel{<}{_{\sim}}$\fi}
\def\gapp{\ifmmode\stackrel{>}{_{\sim}}\else$\stackrel{<}{_{\sim}}$\fi}
\def\mcom{M_{\rm COM}}
\def\mpsr{M_{\rm PSR}}

\def\dpsr{d_{\rm PSR}}
\def\rcom{R_{\rm COM}}
\def\rrl{R_{\rm RL}}
\def\Uc{\rm m_{F390W}}
\def\Vc{\rm m_{F606W}}
\def\Ic{\rm m_{F814W}}
\def\Hc{\rm m_{F656N}}

\def\comC{COM-M5C}
\def\psrC{PSR J1518$+$0204C}
\newcommand{\msp}{J1518$+$0204C}

\def\Msun{M_{\odot}}
\def\msun{$M_{\odot}$}

\def\Lsun{L_{\odot}}
\def\Rsun{R_{\odot}}

\newdimen\minuswidth    

\setbox0=\hbox{$-$}

\minuswidth=\wd0

\catcode`@=\active

\def@{\kern\minuswidth}

\setbox0=\hbox{\rm0}

\shorttitle{}

\shortauthors{}

\begin{document} 

\title{Radio timing and optical photometry of the black widow system PSR J1518+0204C in the globular cluster M5\footnote{Based on observations
    collected with the NASA/ESA HST (Prop. 19835), obtained at the
    Space Telescope Science Institute, which is operated by AURA,
    Inc., under NASA contract NAS5-26555.} }

\author{
C. Pallanca\altaffilmark{1}, S. M. Ransom\altaffilmark{2}, F. R. Ferraro\altaffilmark{1}, E. Dalessandro\altaffilmark{1}, B. Lanzoni\altaffilmark{1}, J. W. T. Hessels\altaffilmark{3,4}, I. Stairs\altaffilmark{5}, P. C. C. Freire\altaffilmark{6}}

\affil{\altaffilmark{1} Dipartimento di Fisica e Astronomia, Universit\`a di Bologna, Viale Berti Pichat 6/2, I-40127 Bologna, Italy }

\affil{\altaffilmark{2} National Radio Astronomy Observatory (NRAO), 520 Edgemont Road, Charlottesville,
Virginia 22901, USA.}
 
\affil{\altaffilmark{3}  ASTRON, the Netherlands Institute for Radio Astronomy, Postbus 2, 7990 AA, Dwingeloo, The Netherlands}

\affil{\altaffilmark{4}  Anton Pannekoek Institute for Astronomy, University of Amsterdam, Science Park 904, 1098 XH Amsterdam, The Netherlands}

\affil{\altaffilmark{5} Department of Physics and Astronomy, University of British Columbia, 6224 Agricultural Road Vancouver, BC V6T1Z1, Canada}

\affil{\altaffilmark{6} Max-Planck-Institute f$\ddot{u}$r Radioastronomie, D-53121 Bonn, Germany}

\date{04 September, 2014}

\begin{abstract}

We report on the determination of astrometric, spin and orbital parameters for \psrC , a 
``black widow'' binary millisecond pulsar  in the globular cluster M5.
The accurate  position
and orbital parameters
obtained from  radio timing allowed us to search for the optical companion.
By using WFC3/HST  images we identified a very faint variable star ($\Uc \gapp 24.8$, $\Vc \gapp 24.3$, $\Ic \gapp  23.1$) located at only 0.25\arcsec\ from the  pulsar's timing position.
Due to its strong variability, this star  is visible only in a sub-sample of images. However, the light curve obtained folding the available data with the orbital parameters of the pulsar shows a maximum at the pulsar inferior conjunction and a possible minimum at the pulsar superior conjunction. 
Furthermore, the shape of the optical modulation indicates a heating process possibly due to the pulsar wind. This is the first identification of an optical companion to a black widow pulsar  in the dense stellar environment of a globular cluster.

\end{abstract} 


\section{INTRODUCTION}\label{intro}
According to the canonical formation scenario, millisecond pulsars (MSPs) form in binary systems containing a neutron star (NS) eventually spun up  to millisecond periods by  mass accretion from an evolving companion, that, in turn, is expected to become a white dwarf \citep[WD; e.g][]{lyne87, alpar82, bhattacharya91}.
About 40\% of known MSPs are found in globular clusters\footnote{see {\it http://www.naic.edu/$\sim$pfreire/GCpsr.html} for an updated list} (GCs), 
although the Galaxy is 100 times more massive \citep[$2.4\times10^{11}\le M/ \Msun \le 1.2\times 10^{12}$;][]{little87,kochanek96} than the entire GC system.
This 
 is partially caused by selection effects, since GCs have been searched very deeply by radio surveys.
Despite this, the over-abundance of MSPs in GCs compared to the Galactic disk is real and very large, and provides
a strong indication that dynamical  interactions greatly enhance the formation of these objects. 
In fact, in the Galactic field the only viable formation channel for MSPs is the evolution of primordial binaries, while in GCs dynamical interactions can promote the formation of binaries suitable for recycling NSs into MSPs \citep[e.g.,][]{ davies98}.
In particular, the ultra-dense cores of GCs are very efficient  ``factories" for generating exotic objects, such as low-mass X-ray binaries, cataclysmic variables, blue stragglers and MSPs \citep[e.g.][]{bailyn95, verbunt97,ferraro01a}. 
Indeed, these objects are thought to result from the evolution of various kinds of binary systems originated and/or hardened by stellar interactions \citep[e.g.][]{clark75, hillsday76, bailyn92, ivanova08}, and are therefore considered as powerful diagnostics of GC dynamical evolution \citep[e.g.][]{fe95, goodmanhut89,hut92, meylanheggie97, pooley03, fregeau08, ferraro09, fe12}. 

Studying the optical emission properties of binary MSP companions is important to better constrain the orbital parameters and to clarify the evolutionary status of these systems.
In GCs, it also represents a crucial tool for quantifying the occurrence of dynamical interactions, understanding the effects of crowded stellar environments on the
evolution of binaries, determining the shape of the GC potential well, and estimating the mass-to-light ratio in the GC cores \citep[e.g.,][]{phinney92, bellazzini95, possenti03, ferraro03a}.
Despite their importance,  only  eight MSP companions in six GCs have been identified so far  \citep[][]{edmonds01, ferraro01a, edmonds02, sigurdsson03, ferraro03b, bassa03, cocozza08, pallanca10, pallanca13comM28I}.  Three of them are likely helium WDs, in agreement with the expectations of the MSP recycling scenario, while the other five are non-degenerate stars, which are  thought to be either the result of  a different evolutionary path or the product of an exchange interaction. 

``Black Widows'' (BWs) are MSPs characterized by  an unmeasurably small eccentricity and a very small mass function \citep[thus indicating a companion mass  smaller than $0.05\Msun$;][]{roberts13}.  
In most cases these pulsars show eclipses in the radio signal suggesting that the companion is a non-degenerate, possibly bloated star. In particular, in some systems the radio eclipse lasts for a significant fraction of the orbit,  implying that the eclipsing region at the position of the companion is larger than its Roche lobe (RL). This suggests that the obscuring material is the plasma released by the companion because of the energy injected by the pulsar.  

 After the discovery of the first BWs in GCs, \citet[][]{king03} observed that they represented
a much larger fraction of the MSP population than in the Galactic disk. Because of this,
they proposed that BWs form much more often than other types of MSPs in GCs, going even as far as to suggest
that perhaps they form {\em exclusively} in GCs and that the BWs in the Galactic disk were formed in GCs and later
ejected. 
However, in the last years the number of BWs discovered in the Galactic field has significantly increased, both in blind surveys \citep{burgay06,bates11,keith12}
and particularly in the surveys targeted at Fermi unidentified sources
\citep[e.g.,][]{ransom11,keith11}
suggesting that BWs
can  form directly in binary systems in the Galactic field,
with no need for an exchange interaction.
If this view is correct, then the percentages of BWs as a fraction of the
total MSP population should be similar in the Galaxy and in GCs,
particularly the GCs with a low interaction rate {\em per binary}
\citet{verbuntfreire14}: In both environments, an LMXB, once formed, can evolve
undisturbed towards the MSP binary stage. Some BWs should survive
even in GCs with a high interaction rate per binary, because they have very small
orbits that are unlikely to be disrupted.
Hence studying  the numbers and properties of BWs both in the Galactic field and in GCs is
important to test these two hypotheses (formation through dynamical interactions or
the evolution of primordial binaries). This is important because
their formation is still very poorly understood.

Up to now, only a few companions to BWs have been detected   \citep[][]{fruchter88b, stappers96b,  stappers99, stappers00,reynolds07, vankerk11, pallanca12, romani12, breton13},   none of them in a GC.
The companions to BWs in the Galactic field are found to be  low-mass objects likely ablated and with a  partially filled RL.
Moreover, these objects show an IR/optical/UV modulation of a few magnitudes correlated with the orbital motion.
The light curve of BW companions is usually characterized by a  maximum around pulsar inferior conjunction ($\Phi\sim0.75$) and one minimum around pulsar superior conjunction ($\Phi\sim0.25$).
This shape is thought to be due to the pulsar wind  heating the side of the companion that faces the pulsar.
  
Here we present  the  first identification of an optical companion to a BW pulsar in the GC  M5. 
M5 (NGC 5904) is a dynamically evolved GC \citep{fe12} with intermediate central density and concentration (\citealp[$\log \rho_0=4.0$ in units of $M_\odot/$pc$^3$;][]{pryormeylan93}; \citealp[c=1.66;][]{miocchi13}) and relatively high metallicity \citep[{[Fe/H]} $\sim$ -1.3,][]{carretta09}   located at $\sim 7.5$ kpc from the Earth \citep[][ 2010  version]{ferraro99, harris96}. M5 harbors  five MSPs
\citep{anderson97,hessels07,freire08}. Among them, \psrC\ deserves special attention since it is a BW system.
This pulsar    has a spin period of 2.48 ms and it is in a 2.1-hr orbit with a companion of minimum mass  $\sim0.04 \Msun$. It shows regular eclipses for $15\%$ of its orbit, as well as eclipse delays at eclipse ingress and egress, which can last up to 0.2 ms, and are presumably due to dispersive delays as the pulsar passes through the ionized wind of its companion \citep[][]{hessels07}. 
If BWs are  directly created in binary systems without the need for exchange
  interactions, then \psrC\ should resemble the BWs in the Galactic field. In fact,
  the very short orbital period ($\sim2$ hr) and M5's very low interaction
  rate per binary both suggest that it unlikely that \psrC\
  was significantly disturbed following the LMXB stage.
  
In Sect. 2 the results of the radio timing are reported, while the optical observations and the identification of the companion to the pulsar are described in Sect. 3. The results are discussed in Sect. 4. In Sect. 5 we  present some concluding remarks with attention to possible future studies.

\section{RADIO TIMING}

\subsection{Observations and data processing}

PSR \msp\ 
was discovered as part
of a series of deep 1.4-GHz observations of several GCs
with the Arecibo telescope in the summers of 2001 and 2002.  Details
of the search observations, methodology, and follow-up timing
measurements of the new pulsars are described in detail by
\citet{hessels07}, and so we only briefly discuss the timing
observations and methodology for PSR \msp\ here.

All observations were made using one to four Wideband Arecibo Pulsar
Processors  \citep[WAPPs;][]{wapps},  each of which provides 100\,MHz of
bandwidth.  We took observations in search mode, resulting in data
with 256 frequency channels per WAPP, and a time resolution of
64\,$\mu$s.  In order to avoid known sources of radio frequency
interference (aka RFI), one WAPP was centered near 1170\,MHz and the
remaining 2$-$3 WAPPs were placed such that together they covered a
continuous frequency band of 200 or 300\,MHz centered near 1470\,MHz. 

In the early years (2002-2004), we were taking many observations scattered through the year in order to better determine the astrometric and spin parameters for all the pulsars in the cluster. Since then, we have prioritized 1-week campaigns in order to improve the measurement of the rate of advance of periastron of M5B, as detailed in Freire et al. (2008), with a few scattered observations that helped improve the proper motions of all the pulsars. Given the short orbital period of M5C, we get excellent, but quasi-random orbital coverage of this pulsar, since the observing strategy has been mostly designed with M5B in mind.

The data were processed using standard techniques \citep[e.g.][]{lk12}
with tools in the {\tt
  PRESTO}\footnote{\url{http://www.cv.nrao.edu/~sransom/presto}}
software suite.  We folded the lower and contiguous upper frequency
bands modulo the predicted pulse period using {\tt prepfold}, removed
strong narrow-band or transient broad-band RFI as identified by eye
from the folds, then averaged the high signal-to-noise detections
together to make a template pulse profile.  We determined times of
arrival (TOAs) every 3$-$15 minutes (based on the strength of the
detection, primarily due to diffractive scintillation) by
cross-correlating the template profile with the folded data using the
 routine {\tt get\_TOAs.py}\footnote{The {\tt get\_TOAs.py} routine is based on the algorithm  described by \citet{taylor92}.},   resulting in 2078 TOAs covering a period of
about 9 years.

We determined a phase-connected timing solution using {\tt
  TEMPO}\footnote{\url{http://tempo.sourceforge.net}}, by fitting for
astrometric, spin, and orbital parameters, along with a single average
 dispersion measure.  We corrected arrival times to the TT(BIPM) time
standard and used the DE421 solar system ephemerides \citep{folkner08}. 
Only 1398 TOAs
were actually used in the fit as we ignored those TOAs between orbital
phases of 0.2$-$0.38 due to the regular pulsar eclipse and the pulse
delays during eclipse egress which would systematically skew orbital
parameter fits.  The eccentricity of the  2.08-hr orbit was too small
to measure and we therefore fixed it, as well as the argument of periapsis
($\omega$), both to zero.

\begin{figure*}[t]
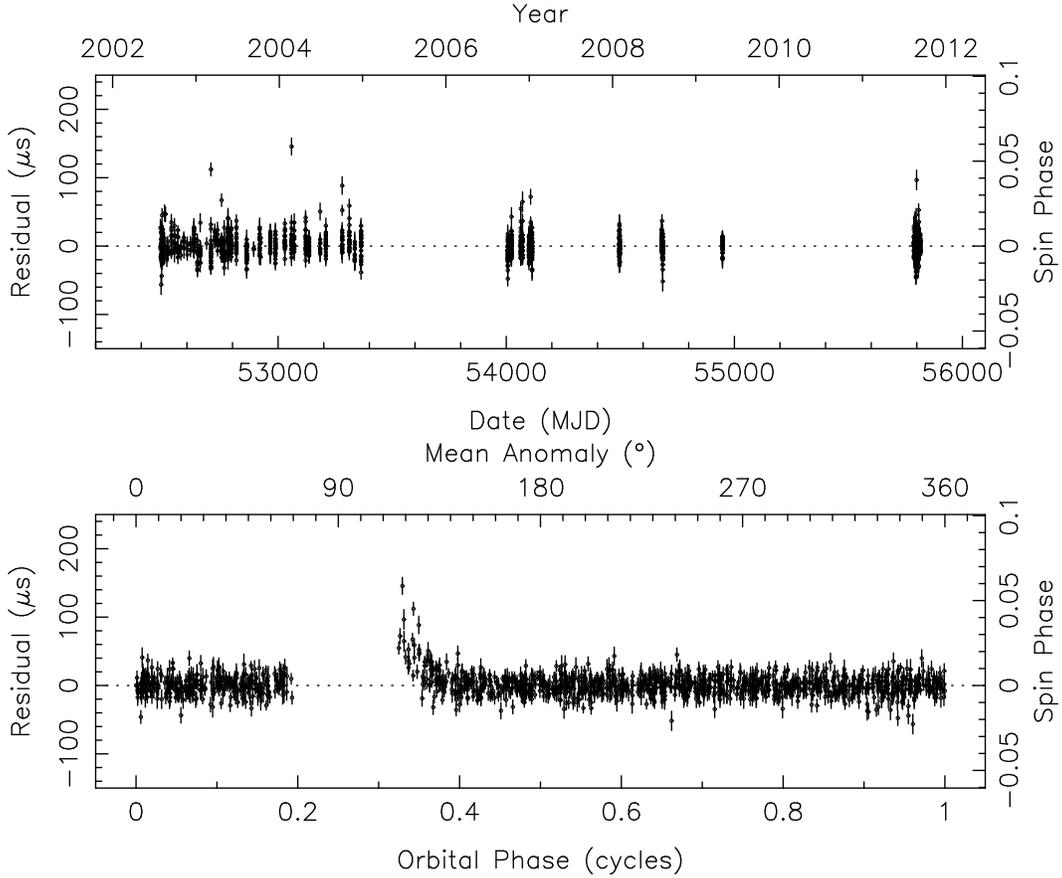

\begin{center}
\includegraphics[angle=270, width=140mm]{fig1a.ps}
\includegraphics[angle=270, width=140mm]{fig1b.ps}
  \caption[Timing residuals for MSP \msp.]{Timing residuals for PSR  \msp.  The top panel shows the
   post-fit timing residuals of the timing solution described in
   Table~\ref{Tab:timing} as a function of date (in MJD).  The
   weighted RMS of the residuals is 12.3\,$\mu$s based on1398 arrival
   times.  The bottom panel shows post-fit residuals as a function of
   the orbital phase (mean anomaly), where the radio eclipse is
   obvious near orbital phase 0.25 (pulsar superior conjunction).  
   Arrival times between the
orbital phases of 0.2 and 0.38 are not among the 1398 used in the timing fit, but are displayed here to illustrate the extra dispersive delay near superior conjunction, which is due to plasma emanating from the companion star.}\label{M5Cfig:radiores}
\end{center}
\end{figure*}

The final timing solution, reported in Table~\ref{Tab:timing} and
shown in Figure~\ref{M5Cfig:radiores}, had weighted RMS residuals of
12.3\,$\mu$s and a reduced-$\chi^2$ of 1.88 with 1385
degrees-of-freedom.  We estimated the errors on the fit parameters
using a bootstrap error analysis with 4096 iterations.  

\subsection{Results}

The timing solution includes a very precise position and
proper motion, which are absolutely required for
follow-up of the object in the crowded field of M5.
The system is located at
$\alpha {\rm (J2000)} =  15^{\rm h} 18^{\rm m} 32\fs788893(21)$ and $\delta {\rm  (J2000)}= 02\degr 04 \arcmin 47\farcs8153(8)$,
which is only $7.5\arcsec$ (corresponding to $\sim0.27$ core radii; $r_c=28\arcsec$) from the nominal cluster center as determined by \citet{miocchi13}.
This is well within the core radius of the cluster.
The proper motion is entirely consistent with
the previously measured proper motions of PSR B1516+02A and B
\citep{freire08}, indicating that it is mostly due to
the motion of the cluster as a whole. The peculiar velocities
of these pulsars within the cluster are still too
small to be measured with our current timing precision.

The observed spin period derivative of the pulsar is a combination of
several contributions, the main ones being
due to its intrinsic spin period derivative, $\dot{P}_{\rm int}$, the
acceleration of the pulsar in the gravitational field
of the cluster projected along the line of sight $a_l$ and the
Shklovskii's effect, due to the pulsar's total proper motion $\mu$:
\begin{equation}
\left( \frac{\dot{P}}{P} \right)_{\rm obs} = \left( \frac{\dot{P}}{P}
\right)_{\rm int} + \frac{a_l}{c} + \frac{\mu^2 d}{c},
\end{equation}
where $d$ is the distance to the cluster, 7.5 kpc. The Galactic
acceleration is generally
small or of the same order as the last term, which is the smallest.

Using a simple empirical model of GC interiors \citep{king62,freire05}
and the cluster's central velocity dispersion of 5.5 km s$^{-1}$
\citep[][2010 version]{harris96}, we estimate that the
maximum cluster acceleration along the line of sight at the location
of M5C is $a_{\rm max} = \pm 1.54 \times 10^{-9} \rm m s^{-2}$.
This would need to be of the order of $\pm 2.6 \times 10^{-9} \rm m
s^{-2}$ to fully account for the observed $\dot{P}$
already without the contribution from the Shklovskii effect (which
amounts to $+0.5 \times 10^{-9} \rm m s^{-2}$).
From this, we can infer minimum and maximum values for $\dot{P}_{\rm
int}$ of $0.9$ and $3.5 \times 10^{-20}$, which implies
a magnetic field $ 1.5 \times < B_0 < 3.0 \times 10^8$ G and a
characteristic age
$1.1 < \tau_c < 4.3$ Gyr. These values are typical among the Galactic
MSP population.

We also detect a variation of orbital period with very high significance;
its time derivative $\dot P_{orb}$ is $-0.914(23)\times10^{-12}$.
Considering the typical masses of such binary systems the measured value is one order of magnitude larger than 
the expected contribution due to the emission of gravitational waves predicted by general relativity.
Such a discrepancy is a common feature for BW systems
\citep{lazaridis11}, where significant changes of the orbit are most likely caused by tidal dissipation leading to changes in the gravitational quadrupole moment of the companion \citep[which are not negligible in the case of a distorted companion, or in the presence of an interaction between the companion and the pulsar;][and references therein]{handbook}. The effect is {\em not} due to the acceleration of the pulsar in the cluster:
even if it were accelerating with $\pm a_{\rm max}$, the contribution to the orbital variation would only be $\dot{P}_{orb, k} =  \pm 0.04 \times 10^{-12}$.

We also detect a variation of the apparent slowdown of the pulsar,
$\ddot{\nu}$, although with much lower significance. This is likely caused
by a small change in the gravitational acceleration of the cluster
and nearby stars at the position of the pulsar, something to be
expected considering that the pulsar is, at least in projection,
very near the center of the cluster, where the stellar density is highest.

\section{OPTICAL PHOTOMETRY OF THE COMPANION STAR}\label{identification}
\subsection{Observations and data analysis}\label{Sec:dataan}
By taking advantage of the  precise position of \psrC\ obtained from the radio timing, we used high resolution  Hubble space telescope (HST) images to search for the optical companion to this  pulsar.
The photometric data-set  consists of  a set of images obtained with the ultraviolet-visible (UVIS) channel of the WFC3.  
The WFC3 UVIS CCD consists of two twin detectors with a pixel-scale of $\sim 0.04\arcsec$/pixel and a global field of view (FOV) of $\sim 162\arcsec\times 162\arcsec$.  
Our analysis has been focused only on  CHIP1, which contains the  pulsar region. 

The analyzed images have been acquired through four filters, in two different epochs.
The first epoch images have been obtained on 2010 July 5 (Prop. 11615, P.I.  Ferraro).
The data-set consists of: 6 images  in the F390W filter  with an exposure time $t_{\rm exp}=500$ s each, 4 images in F606W  with $t_{\rm exp}=150$ s, 4 images in F814W  with $t_{\rm exp}=150$ s, and 6 images in F656N (a narrow filter corresponding to H$\alpha$) with exposure times ranging from $t_{\rm exp}=800$ s to $t_{\rm exp}=1100$ s.  
The second epoch images have been obtained during four visits between 2012  June 6 and 2012 June 9 (Prop. 12517, P.I. Ferraro), using   the same three wide filters  as the first epoch. 
In particular, the data-set consists of: 4 images obtained through the F390W filter  with an exposure time  $t_{\rm exp}=735$ s each, 8 images in F606W  with $t_{\rm exp}=350$ s, and 12 images in F814W  with $t_{\rm exp}=230$  s and $240$ s.

The photometric analysis  has been performed on the WFC3 ``flat fielded" (flt) images,  once corrected\footnote{For more details on the applied correction and on the Pixel Area Map definition see the WFC3 Data Handbook.} for  ``Pixel-Area-Map'' (PAM) by using standard IRAF procedures. The photometric analysis has been carried out by using the {\sc daophot} package \citep[][]{stetson87}. For each image we modeled the point spread function (PSF) by using a large number ($\sim 100$) of bright and nearly isolated stars.
The PSF model and its parameters have been chosen by using the  {\sc daophot} {\tt PSF} routine  on the basis  of a Chi-squared test.  A Moffat function \citep[][]{moffat69} provides the best fit to the data in  all cases.
 All F390W, F606W and F814W images have been combined with {\sc daophot} {\tt MONTAGE2} in order to produce a master frame.
On  this combined image we then performed  a source detection analysis by using {\sc daophot} {\tt FIND} and a  $3\sigma$ detection limit, where $\sigma$ is the standard deviation of the measured background. 
 Finally, using the star list thus obtained as a reference master list, we performed the PSF-fitting  in each image by using the {\sc daophot} packages {\tt ALLSTAR} and {\tt ALLFRAME} \citep{stetson87, stetson94}.
For each star the magnitudes estimated in different images have been homogenized \citep[see][]{ferraro91,ferraro92} and
their weighted mean and standard deviation have been finally adopted as the star mean magnitude and its photometric error.
In the final catalog we reported both single image magnitudes and the mean magnitudes in each filter.

Since the FOV of the WFC3 images suffers heavily  from geometric distortions, we corrected the instrumental positions of stars by applying the equations reported by \citet[][]{bellini09} and \citet{bellini11}. 
We then transformed  the  WFC3 catalog to the absolute astrometric system ($\alpha$, $\delta$) by using the stars in common with the HST WFPC2 catalog from \citet[][]{lanzoni07} as secondary astrometric standards, by using the cross-correlation software CataXcorr\footnote{CataXcorr is a code aimed at cross-correlating
catalogs and finding astrometric solutions, developed by
P. Montegriffo at INAF - Osservatorio Astronomico di Bologna.  This
package, available at {\it http://davide2.bo.astro.it/$\sim$paolo/Main/CataPack.html}, has been successfully used in a large number of papers by our group in the
past 10 years.}. The 
astrometric solution has an accuracy of $\sim0.2''$ in both $\alpha$ and $\delta$.

Finally, the WFC3 instrumental magnitudes have been calibrated to the VEGAMAG system by using the photometric zero-points and the procedures reported on the WFC3 web page.\footnote{http://www.stsci.edu/hst/wfc3/phot\_zp\_lbn}

\subsection{The companion to \psrC}
\begin{figure*}[t]
\begin{center}
\includegraphics[width=140mm]{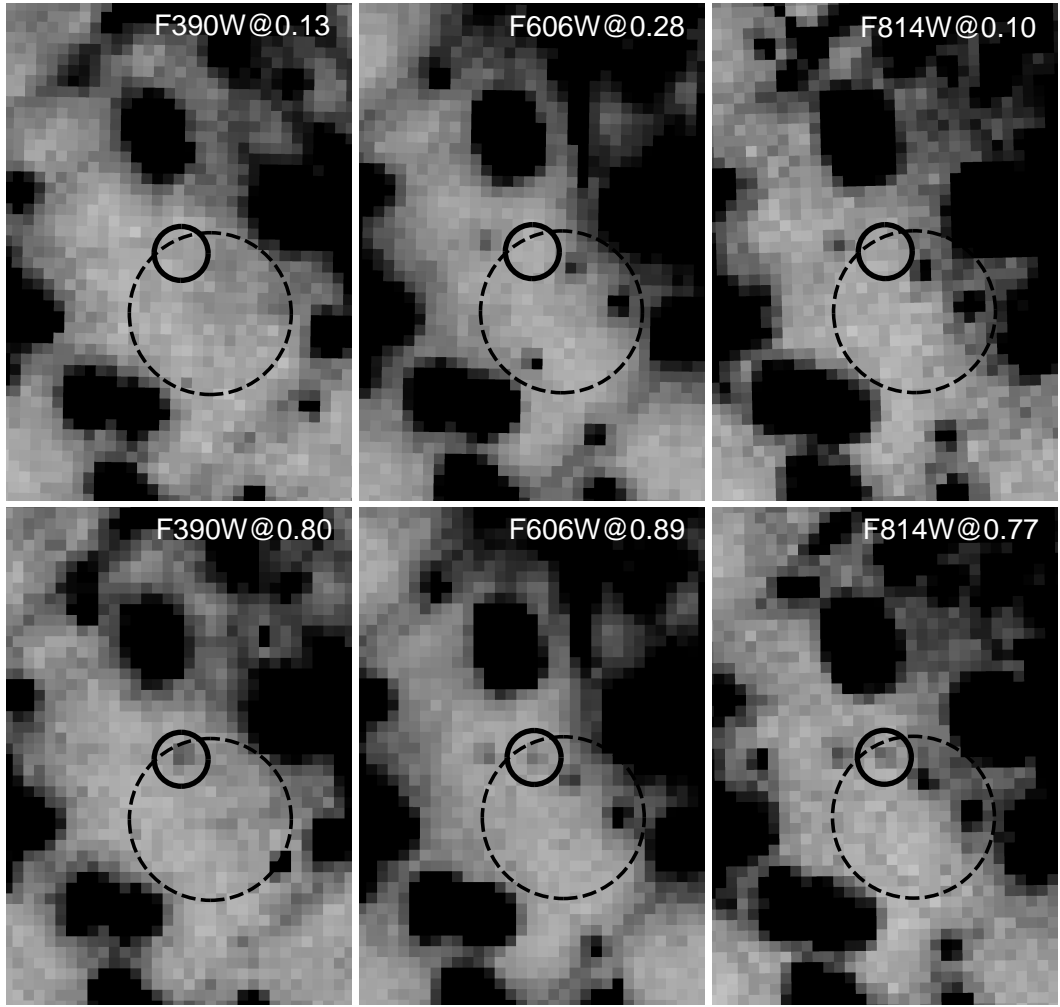}
  \caption[HST images of the region around \psrC.]{HST images of the  $1.3 \arcsec\times 1.8 \arcsec$ region around the nominal position of \psrC. The filters and orbital phases are labelled in each panel.  The dashed circle of radius $0.3\arcsec$ (of the order of the optical astrometric uncertainty)  is centered on the radio-band pulsar position. The solid circles indicate the identified companion star. It is clearly visible in the lower panels (at about the pulsar inferior conjunction), while it disappears in the upper panels (at about the pulsar superior conjunction). 
  }\label{M5Cfig:map}
\end{center}
\end{figure*}

In order to  identify the companion to \psrC\ we searched for peculiar objects located within $1\arcsec$ from the nominal pulsar position  (see Table \ref{Tab:timing}). 
At a first visual inspection of the pulsar region, it was possible to identify a star showing strong variability and  lying at $\sim 0.25\arcsec$  from the pulsar, which is within the optical astrometric uncertainty.
This star is  very faint and  visible only  in a few images, while it is  below the detection limit in the other frames (see Figure \ref{M5Cfig:map}).

 For this reason we performed a visual inspection of all the 44 available images, whenever possible forcing the fitting procedure to determine the magnitude of this faint star.
We were able to measure its magnitude in only 14 images (4 in the F390W, 3 in the F606W and 7 in the F814W filters) 
however this limited data set was sufficient to clearly assess the variability of this star, which showed  quite significant luminosity variations: 
 $\Delta \Uc \sim 0.32$ mag (from $\Uc=24.83 \pm 0.17$ to $\Uc=25.15 \pm 0.22$),  $\Delta\Vc \sim 0.62$ mag (from $\Vc=24.34 \pm 0.20$ to $\Vc=24.96 \pm 0.19$), and $\Delta\Ic \sim 0.87$ mag (from $\Ic=23.13\pm 0.07$ to $\Ic=24.00 \pm 0.28$).
Unfortunately, the object is under the detection threshold in all F656N images, hence we do not have any $\Hc$ measure. 
In the other images this star is not detected, most likely because its flux is below the  detection threshold. For each band, we estimated the detection threshold  as the average value of the magnitudes of the five faintest detected stars within 20\arcsec\ from the pulsar position, obtaining ${\rm DT_{F390W}} \sim26.59\pm 0.13$, ${\rm DT_{F606W}}\sim25.67 \pm 0.10$ and ${\rm DT_{F814W}}\sim24.41 \pm 0.14$.
In turn, these values imply the following lower limits to the amplitudes of variation: $\Delta\Uc \ge 1.76$, $\Delta\Vc \ge 1.33$ and $\Delta\Uc \ge 1.28$.
Moreover
this star shows  a magnitude scatter larger than that computed for objects of similar luminosity. 

\begin{figure*}[b]
\begin{center}
\includegraphics[width=140mm]{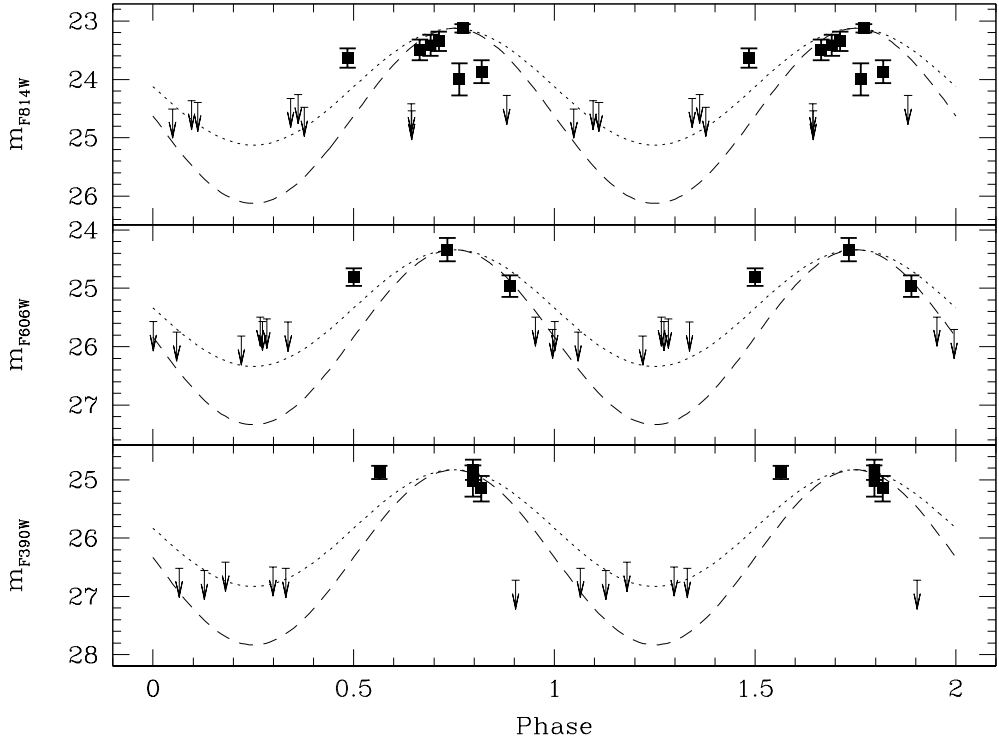}
  \caption[Light curve of the companion to \psrC.]{The observed light curve of the companion to \psrC\ folded with radio timing orbital parameters. The arrows are the estimated  upper-limits to the magnitude for the images where the star is below the detection threshold. 
  The dotted and dashed   lines are sinusoidal  first-guess models of the light curve, with amplitudes of two and three magnitudes, respectively.   Note that the latter model is  in agreement with most of the upper limits. 
   However,  the light curve does not look as a perfect sinusoid, in agreement with the models of \citet{breton13} calculated for similar objects.
  }\label{M5Cfig:lc}
\end{center}
\end{figure*}

In order to establish  whether  the  magnitude of variation is related to the pulsar's  orbital phase (and hence establish  a physical  connection between this star and the pulsar), we computed the light curve in the three bands  folding each measurement (using  a magnitude upper limit  for the images where the star is not detected) with the orbital period and the  ascending node of the pulsar  (see Table \ref{Tab:timing}). 
Although the available data do not allow a complete coverage of the orbital period, the  flux modulation of the star  nicely correlates with the pulsar's orbital phase  (see Figure \ref{M5Cfig:lc}). In fact, the data are consistent with a luminosity maximum (in each band) around $\Phi=0.75$,  corresponding to the pulsar inferior conjunction (when we are observing the companion side facing the pulsar) and a luminosity minimum (at least a few magnitudes fainter) around $\Phi=0.25$, corresponding 
to the pulsar superior conjunction (when we are observing the back side of the companion).
Such a  shape is the typical light curve expected when the surface of the companion is heated by the pulsar flux.

For reference, we over-plotted   sinusoidal functions\footnote{Note that the light curve can be more complicated
  than a sinusoid, as found for a few Galactic field BWs by  \citet{breton13}. }   with a maximum at $\Phi=0.75$ and a minimum at $\Phi=0.25$ onto the observed light curve. The first important point to note is that in order to account for most of the upper limits where the star is not detected, an amplitude variation of  about three magnitudes is required (see  Figure \ref{M5Cfig:lc}).
Such a large modulation ($\Delta$mag$\sim3$ mags) is in good agreement with what  is observed for similar objects  in the Galactic field \citep[e.g.][]{stappers99, stappers01, pallanca12}.
Second, despite the low significance of the detection, there are some hints that the light curve could be asymmetric (e.g. the decrease to minimum seems to be steeper than the increase to maximum),  possibly due to an asynchronously rotating companion as in the case of PSR J2051$-$0827 \citep[][]{doroshenko01,  stappers01}.

As evident from Figure \ref{M5Cfig:map}  two other stars are located within the optical astrometric uncertainties. In the color magnitude diagrams (CMDs) they are both located in the lowest part of the main sequence and from a variability analysis their light  curves appear to be  dominated by the large photometric errors. In particular they seem to have different behaviors in the various filters and they do not show any correlation with the orbital motion of the system. Hence, we can safely rule out these objects as possible candidate companions.

All these pieces of evidence suggest that the identified variable star is the companion to \psrC\ and  we name it \comC.
Since the available $\Uc$, $\Vc$ and $\Ic$  measurements do not allow 
us to have  reliable measures of the mean magnitudes of this star, 
 in the following analysis we will use the values at maxima ($\Uc=24.83$,  $\Vc=24.34$ and $\Ic=23.13$) and a plausible range of magnitude variation ($\sim 3$ mags). 

Figure \ref{M5Cfig:cmd} shows the position of \comC\  in the ($\Vc$, $\Vc - \Ic$) and in the  ($\Uc$, $\Uc - \Vc$)   CMDs.
As can be seen, \comC\ is located at faint magnitudes between the  main sequence (MS) and the WD cooling sequence,  thus suggesting that it is probably a non-degenerate or a semi-degenerate, low mass, swollen star. Indeed similar objects have been previously identified in Galactic GCs \citep[see][] {ferraro01com6397, edmonds02, cocozza08, 
pallanca10, pallanca13comM28I}.

\section{DISCUSSION}
\begin{figure*}[p]
\begin{center}
\includegraphics[width=140mm]{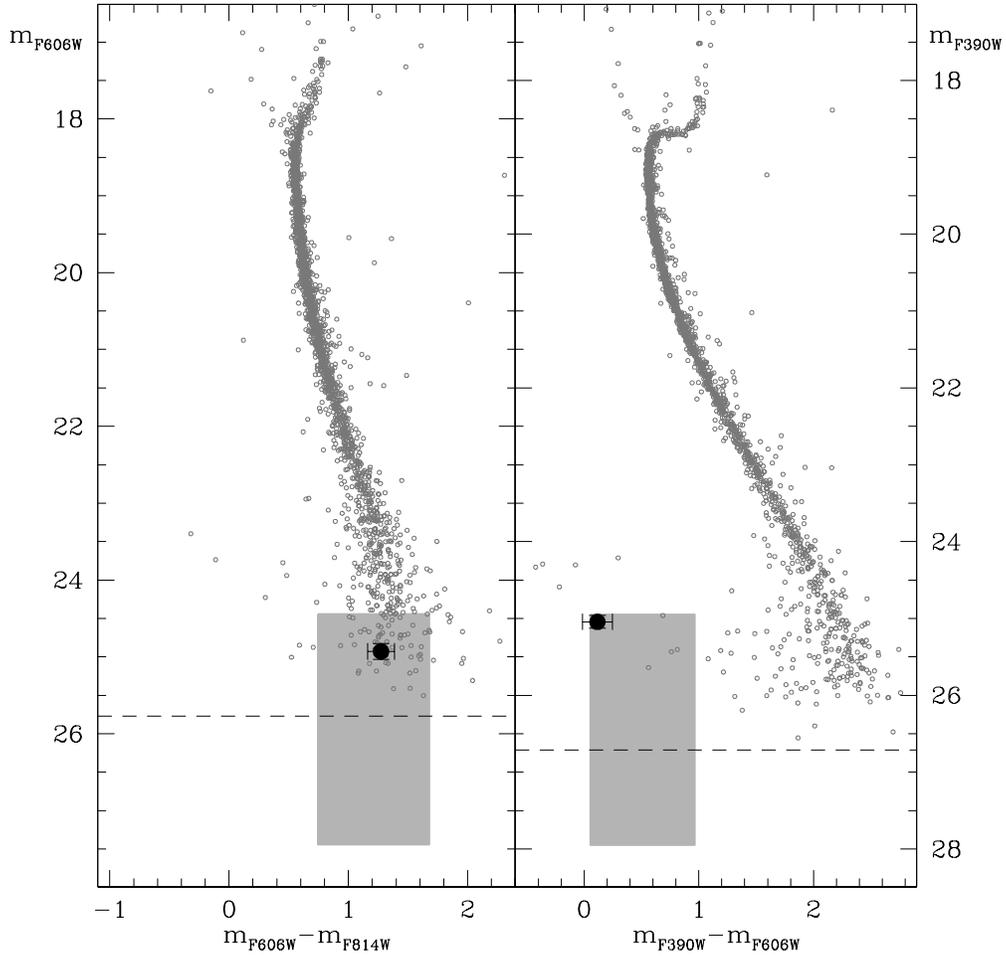}
  \caption[CMD position of the companion to \psrC.]{Optical CMDs defined by  all stars detected within $10\arcsec$ from the pulsar nominal position (gray circles). 
  The dashed lines correspond to the detection limits in the F606W and F390W bands (left and right panel, respectively).
  In each panel the black dot marks the   position  of \comC\ as obtained from the average of the measured values. However, such a location is biased by the under-sampling of the light curve.
   Hence, by taking into account the magnitude modulation we report the possible location of the companion star. 
   In particular,  the magnitude ranges between the measured maximum and a minimum assumed to be three magnitudes fainter,  while the color has been calculated as the color at maximum, with an uncertainty estimated to be of the order of the standard deviation shown by  objects with similar magnitudes.
    Therefore,  \comC\ turns out to be located between the MS and the WD  loci.
     }\label{M5Cfig:cmd}
  \end{center}
\end{figure*}

We have constrained the mass, luminosity and temperature of \comC\ by comparing  its position in the ($\Vc, \Vc-\Ic$) CMD (Figure \ref{M5Cfig:cmd}) with a reference isochrone well reproducing the main evolutionary sequences of MS  stars \citep{girardi00,marigo08}.
We adopted a metallicity [Fe/H] $\sim$ -1.3 \citep{carretta09}, an age  $t=13$ Gyr, a distance modulus $(m-M)_0$=14.37 and a color excess $E(B-V)=0.03$ \citep{ferraro99}. 
In particular,  for the  magnitude of the  companion  we assumed the value at maximum ($\Vc=24.34$) as an upper limit to the luminosity and for its color we adopted the value at maximum  ($\Vc - \Ic = 1.21 \pm 0.46$) as a reference.
The color uncertainty has been estimated from the typical photometric error of stars of similar magnitudes (see the gray shaded region in Figure \ref{M5Cfig:cmd}). 
The resulting effective temperature, luminosity  and mass of the  companion are  $3440{\rm K} \lapp T_{\rm eff} \lapp 5250 {\rm K}$,  $L \lapp 1.19\times 10^{-2} \Lsun$ and  $\mcom \lapp 0.2 \Msun$. 
Given that both the luminosity and temperature of the companion are overestimated, the comparison with the adopted theoretical isochrone gives an upper limit to the companion mass.

Note that if we consider the largest derived value of the mass ($\mcom\sim0.2\Msun$)  and we combine it with the pulsar  mass function, 
 we can rule out very small inclination angles (e.g., $i \lapp 10^\circ$ for a $1.4 \Msun$ NS). 
 However, as 
found for other companion stars \citep[e.g.][]{ferraro03rv6397, pallanca10,   mucciarelli13}, masses  of perturbed stars derived from  their position in  the CMD  might be overestimated\footnote{
In fact, from the measured luminosity we can directly derive the mass through comparison with models of unperturbed MS stars, but 
if the star filled the RL  (and thus  no longer follows the hydrostatic  equilibrium law) such a mass value could be overestimated.}.
 On the other hand, if we assume the lower  mass limit for  core hydrogen burning stars ($\mcom = 0.08 \Msun$) we would obtain an upper limit to the inclination angle of the system 
 $i<30^\circ$. However, such inclinations would be in disagreement with the presence of  a radio eclipse. 

Under the assumption that  the optical emission of \comC\ is well reproduced by a blackbody (BB), 
the stellar radius is $R_{\rm BB} \lapp 0.30\ \Rsun$.
However,  companions to BWs are expected to be affected by the tidal distortion exerted by the pulsar and consequently to be swollen up to fill their RL.
Hence, to justify the presence of eclipses of the radio signal, the size of the RL might be a more appropriate value  \citep[e.g., see PSR J2051$-$0827 and PSR J0610$-$2100][]{stappers96b, stappers00, pallanca12}. According to \citet[][]{eggleton83} the RL radius can be computed as: \begin{displaymath} \frac{\rrl}{a} \simeq \frac{0.49q^{\frac{2}{3}}}{0.6q^{\frac{2}{3}}+\ln \left(1+q^{\frac{1}{3}}\right)}  \end{displaymath} where $q$ is the ratio between the companion and the pulsar masses ($\mcom$ and $\mpsr$, respectively) and $a$ is the orbital separation. This relation can be combined with the pulsar mass function $f_{\rm PSR}(i,\mpsr,\mcom)=(\mcom\sin i)^3/(\mcom+\mpsr)^2$. By assuming a  NS mass ranging from $\sim 1.2 \Msun$ to $2.5 \Msun$ \citep[][]{ozel12} and leaving the inclination angle to span the entire range of values (between $1^\circ$ and $90^\circ$), this yields  $\rrl\sim0.13-0.89\ \Rsun$.

Under the assumption that the optical variation shown in Figure \ref{M5Cfig:lc} is mainly due to irradiation from the MSP,  reprocessed by the surface of \comC, we can estimate how the re-processing efficiency depends on the inclination angle and, hence, infer the companion mass. To this end, one can compare the observed flux variation ($\Delta F_{obs}$)  between the maximum ($\Phi=0.75$) and the minimum ($\Phi=0.25$) of the light curve, with the expected flux variation ($\Delta F_{exp}$) computed from the rotational energy loss  rate ($\dot{E}$). 
Unfortunately, $\dot{E}$ is not measurable with the available radio observations,  because the observed period derivative is likely significantly affected by acceleration of the pulsar in the gravitational potential of the cluster. However we took as reference the
value measured for some BWs in the Galactic field,  which typically have $\dot{E}$ values ranging from $10^{34}$ to  $10^{35}$ erg s$^{-1}$ \citep{breton13}. 
Actually, since we do not observe the entire light curve, $\Delta F_{obs}$ can just put a lower limit to the reprocessing efficiency. 

\begin{figure*}[p]
\begin{center}
\includegraphics[width=140mm]{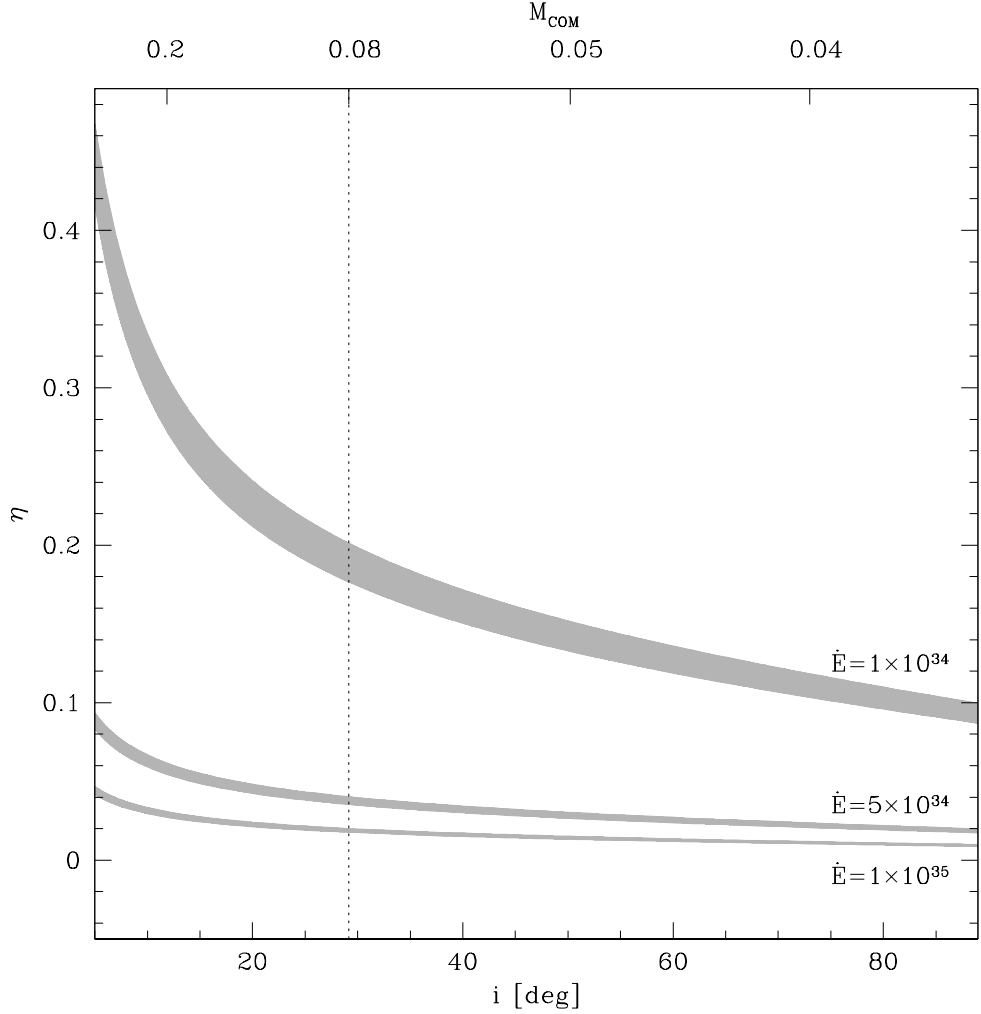}
  \caption[Reprocessing efficiency of the companion to \psrC.]{Lower limit to  the reprocessing efficiency for isotropic emission ($\eta$) calculated as a function of the inclination angle ($i$) and assuming  a magnitude modulation ($\Delta$mag $= 3$ mags).
  The three gray strips correspond to different values of $\dot{E}$: $1.0\times10^{34}$, $5.0\times10^{34}$ and $1.0\times10^{35}$ erg/s,  respectively from top to bottom.
  The thickness of each strip corresponds to a pulsar  mass ranging from $1.24\Msun$ to $2.5\Msun$. 
  On the top axis, the companion masses in units of $\Msun$  (calculated by assuming a $\mpsr=1.4\Msun$) are reported. The dotted line marks the physical limit ($M\gapp0.08\Msun$) for core hydrogen burning stars.}\label{M5Cfig:repr}
\end{center}
\end{figure*}

At first we converted the observed  $\Delta \Vc$ modulation into a flux variation. 
We assumed the maximum measured magnitude ($\Vc =24.34$ at $\Phi=0.75$) and an amplitude of variation $\Delta \Vc =3$ between $\Phi=0.75$ and $\Phi=0.25$, thus obtaining  $\Delta F_{obs} = 2.96 \times 10 ^{-15}$ erg s$^{-1}$ cm$^{-2}$.
On the other hand, the expected flux variation due to irradiation between $\Phi=0.75$ and $\Phi=0.25$ is given by \begin{displaymath} \Delta F_{exp}(i)= \eta \frac{\dot{E} }{a^2}  \rcom^2  \frac{1}{4\pi d^2_{\rm PSR}} \varepsilon (i) \end{displaymath} where $\eta$ is the re-processing efficiency under the assumption of isotropic emission,  $\rcom$ is the radius of the companion star,  which we assumed  to be equal to $\rrl(i)$, $\dpsr$ is the distance of pulsar, adopted to be equal to the GC distance \citep[$\dpsr=7.5$ kpc;][]{harris96,ferraro99} and $\varepsilon (i)$ parametrizes the difference of the re-emitting surface visible to the observer between maximum and minimum\footnote{ In the following we assume 
$\varepsilon (i)=(i/180)(1-\rcom/a)$.
Note that for  $\rcom \ll a$ the second term reduces to zero.
For the two limit configurations we find that in the case of a face-on system ($i=0^{\circ}$), the fraction of the heated surface visible to the observer is constant and hence  no flux variation is expected, while in the case of an edge-on system ($i=90^{\circ}$) the fraction of the heated surface that is visible to the observer varies between 0.5 (for $\Phi=0.75$) to 0 (for $\Phi=0.25$) and hence $\varepsilon=0.5$.}. By assuming $\Delta F_{obs}=\Delta F_{exp}(i)$ between $\Phi=0.75$ and $\Phi=0.25$, we can derive how $\eta$ varies as a function of $i$. The result is shown in Figure \ref{M5Cfig:repr}. 
Obviously, it is important to note that  all the calculations have been performed   assuming a magnitude modulation of about three magnitudes
 and hence in case of a larger amplitude of variation the estimated reprocessing would correspond to a lower limit to the true value.
As an example,  the observed optical modulation can be reproduced considering a system seen at an inclination angle of about $60^{\circ}$,  with a very low mass companion ($\mcom\sim0.04-0.05\Msun$) that has filled its RL, and reprocesses the pulsar flux with an efficiency $\eta \gapp 1-15\%$. Similar results are obtained performing the same calculations in the F390W and F814W bands.
It is important to note that, for simplicity, calculations have been performed assuming a RL filled companion (an upper limit to the true companion size). Consequently,  in the case of a  companion only partially filling its RL the reprocessing efficiency would be larger.

On the other hand, if we use $R_{\rm BB}$ instead of $\rrl$ for the stellar radius, the efficiency increases and for several configurations it becomes larger than 100\%.
In such  cases the only possible scenario would be  an anisotropic pulsar emission. However, given the presence of eclipses and the observed behavior of other similar objects,  $R_{ \rm BB}$ is likely too small to provide a good estimate of  the  companion true physical size.  Future studies are needed to better constrain the system parameters.

\section{SUMMARY AND CONCLUSION}
We obtained a phase-connected radio timing solution of \psrC,  including a very precise position and proper motion. The latter is consistent with the known proper  motion of two other M5 pulsars, indicating that they are all co-moving  because of the cluster motion.
We also estimated the time derivative of the orbital period. As common for BW systems, it is larger than the expected contribution due to the emission of gravitational waves. It is much more likely due to tidal dissipation.
The precise position of the pulsar was used to identify the companion star (\comC) in the optical bands by using high-resolution WFC3/HST images. \comC\ turns out to be a very faint but strongly variable star ($\Delta\Uc \ge 1.76$, $\Delta\Vc \ge 1.33$ and $\Delta\Uc \ge 1.28$). Its location in the CMD in a region between the cluster MS and the WD cooling sequence suggests that it is a low-mass star. This is the first companion star to a BW ever detected in a GC. Because of its faintness the star was detected only in 14 out of the 44 available images, whose phases nearly correspond to the pulsar inferior conjunction, while it is below the detection limit at the pulsar superior conjunction.  
 The analysis of the light curve, obtained adopting a simple sinusoidal model, suggests a variation amplitude of about three mags\footnote{Note that we have considered upper limits estimates when the star was not visible.}, 
 in good agreement with the behavior of  other  BW companions observed  in the Galactic field.
Finally, we estimated a lower limit to the reprocessing factor of the pulsar flux ($\eta \gapp 1-15\%$) by a RL  filling companion.

Unfortunately, the star is too faint and located in  a too crowded region to allow a spectroscopic follow-up with the  available instruments.
However, an optimized photometric follow-up aimed to detect the star at the minimum would give the opportunity to better constrain the light curve shape of the companion and hence to better characterize this binary  system.

One of the main objectives of this study is to contribute with
observational data towards understanding the differences between BW
systems in globular clusters and in the Galaxy. This task is important
because the formation of these systems is still poorly understood, and
it is not known how much it is influenced by the local stellar
density. However, given the limited statistics, particularly of
systems with optical identifications, a detailed comparison of the two
populations is premature at this stage. We refer the reader to a
forthcoming paper with the results of a second optical identification
of a BW in a globular cluster (M. Cadelano et al. 2014; in
preparation).

\section{Acknowledgement}
This research is part of the project COSMIC-LAB funded by the European Research Council 
(under contract ERC-2010-AdG-267675).
The Arecibo Observatory is operated by SRI International
under a cooperative agreement with the National Science Foundation
(AST-1100968), and in alliance with Ana G. M\'endez-
Universidad Metropolitana, and the Universities Space Research
Association.
Pulsar research at UBC is supported by an NSERC Discovery Grant.
J.W.T.H. acknowledges funding from an NWO Vidi fellowship and ERC Starting Grant ``DRAGNET'' (337062). 
 P.C.C.F. gratefully acknowledges
financial support by the European Research Council for the ERC
Starting Grant BEACON under contract 279702.
C.P. thanks the hospitality from NRAO where most of this work has been developed.

\begin{deluxetable}{lc}
\footnotesize
\tablecaption{Parameters for PSR J1518+0204C \label{Tab:timing}}
\tablewidth{0pt}
\tablehead{\colhead{Parameter} & \colhead{Value}}
\startdata
Right Ascension, $\alpha$ (J2000)           \dotfill & $15^{\rm h}\;18^{\rm m}\;32\fs788893(21)$ \\
Declination,     $\delta$ (J2000)           \dotfill & $02\degr\;04\arcmin\;47\farcs8153(8)$ \\
Proper motion in $\alpha$, $\mu_\alpha$(mas yr$^{-1})$ \dotfill & 4.67(14)   \\
Proper motion in $\delta$, $\mu_\delta$(mas yr$^{-1})$ \dotfill &   $-$8.24(36) \\
Dispersion Measure (pc cm$^{-3}$)           \dotfill & 29.3146(6) \\
Span of Timing Data (MJD)                   \dotfill & 52484 - 55815 \\
Number of TOAs                              \dotfill & 1398 \\
Weighted RMS Timing Residual ($\mu$s)       \dotfill & 12 \\
\cutinhead{Spin Parameters}
Pulsar Frequency, $\nu$ (Hz)                  \dotfill & 402.58822840843(9) \\
Frequency Derivative, $\dot\nu$ (Hz/s)           \dotfill & $-$4.2252(2)$\times$10$^{-15}$ \\
Frequency 2nd Derivative, $\ddot\nu$ (Hz/s$^{-2}$) \dotfill & $-$1.45(16)$\times$10$^{-26}$ \\
$P$ Epoch (MJD)                             \dotfill & 52850.0000 \\
\cutinhead{Orbital Parameters}
Orbital Period, $P_{orb}$ (days)            \dotfill & 0.08682882865(3) \\
Projected Semi-Major Axis, $a\sin i$ (lt-s) \dotfill & 0.05732042(57) \\
Eccentricity, $e$                           \dotfill & $0.0000000000$ \\
Epoch of Ascending Node, $T_{asc}$ (MJD)          \dotfill & 52850.00434606(17) \\
Orbital Period Derivative, $\dot P_{orb}$ (s/s)    \dotfill & $-$0.914(23)$\times$10$^{-12}$ \\
\cutinhead{Derived Parameters}
Pulsar Period, $P$ (ms)                     \dotfill & 2.48392757024730(22) \\
Period Derivative, $\dot P$ (s/s)           \dotfill & 2.6055(12)$\times$10$^{-20}$ \\
Mass Function, $f_1$ (\msun)                  \dotfill & 2.68177(14)$\times$10$^{-5}$ \\
Minimum Companion Mass, $m_2$ (\msun)         \dotfill & $\geq$\,0.038 \\
\enddata \tablecomments{Numbers in parentheses represent the formal
errors in the least significant digits as determined by a bootstrap
analysis of the data with 4096 iterations.  The pulsar is assumed to
have a mass of 1.4\,\msun.}  \end{deluxetable}

\end{document}